\documentclass[a4paper, 10pt]{article}
\usepackage[T1]{fontenc}
\usepackage[utf8]{inputenc}
\usepackage{amsfonts}
\usepackage{amssymb}
\usepackage{indentfirst}
\usepackage{amsmath}
\usepackage{amsthm}
\usepackage {geometry}
\usepackage[pdftex]{graphicx}

\newgeometry{tmargin=5.05 cm, bmargin=5.05 cm, lmargin=4.15 cm, rmargin=4.15 cm}

\begin{document}
\setlength{\baselineskip}{13pt}
\newpage
\vfill \vfill \vfill
\begin{center}
\begin{footnotesize}
\textbf{GENERAL RELATIVITY WITH NONZERO COSMOLOGICAL CONSTANT $\Lambda $ AS A GAUGE THEORY}
\end{footnotesize}
\end{center}

\begin{center}
MARTA DUDEK\\
\begin{footnotesize}
\textit{Institute of Mathematics, University of Szczecin}\\
\textit{Wielkopolska 15, 70-451 Szczecin, Poland, EU}\\
\textit{e-mail: marta.dudek@vp.pl}
\end{footnotesize}
\end{center}

\begin{center}
JANUSZ GARECKI\\
\begin{footnotesize}
\textit{Institute of Mathematics and Cosmology Group, University of Szczecin}\\
\textit{Wielkopolska 15, 70-451 Szczecin, Poland, EU}\\
\textit{e-mail: garecki@wmf.univ.szczecin.pl}
\end{footnotesize}
\end{center}

\begin{center}
\begin{footnotesize}
Received 24 July 2017
\end{footnotesize}
\end{center}

\hspace{5 cm}

We show in a new way that the general relativity action (and Lagrangian) in recent Einstein-Palatini formulation is equivalent in four dimensions to the action (and Langrangian) of a gauge field. 

Firstly, we present the Einstein-Palatini (EP) action with cosmological constant $\Lambda \neq 0$ and derive Einstein fields equations from it. Then we consider their action integral in terms of the corrected curvature $\Omega_{cor}$. We will see that in terms of $\Omega_{cor}$ the EP action takes the form typical for a gauge field. Finally, we give a geometrical interpretation of the corrected curvature $\Omega_{cor}$.

This paper is a continuation of the previous paper [17] and it also gives an amended version of the lecture delivered by one of the authors [M.D.] at Hypercomplex Seminar 2017 in Będlewo.

\vspace{0.5 cm}

\section{Einstein-Palatini action for general relativity}

The Einstein-Palatini action with cosmological constant $\widetilde{\Lambda }\neq 0 $ in new formulation [3] reads
\begin{align}
S_{EP}=\frac{1}{4\kappa}\int \limits_{\mathcal{D} }\Big( \vartheta^{i}\wedge \vartheta^{j}\wedge \Omega^{kl}+ \frac{\widetilde{\Lambda }}{6} \vartheta^{i}\wedge \vartheta^{j}\wedge \vartheta^{k}\wedge \vartheta^{l}\Big) \eta_{ijkl}, 
\end{align}
where $\Omega$ is the curvature of the spin connection $\omega $ and $\kappa =8\pi G/c^{4}$. All indices take values $(0, 1, 2, 3)$ and $\mathcal{D}$ means an established 4-dimensional compact domain in spacetime.
$\vartheta^{a}$ denote $1$-forms of the Lorentzian coreper in terms of which the spacetime looks locally Minkowskian, i.e., $g=\eta_{ik} \vartheta^{i}\otimes \vartheta^{k}$, $\eta_{ik}=diag(1,-1,-1,-1)$.\\
$\eta_{ijkl}$ is completely antysymmetric Levi-Civita pseudotensor: $\eta_{0123}=\sqrt{|g|}$, where $g:=det(g_{ik})$. In a Lorentzian coreper $|g|=1$. Spin connection $\omega $ is a general metric connection (or Levi-Civita connection) in Lorentzian coreper. 

For convenience we will write the cosmological constant $\widetilde{\Lambda }\neq 0$ in the form $\widetilde{\Lambda }=\epsilon \Lambda $, where $\Lambda >0$ and $\epsilon =\pm 1$. In conseqence, if $\epsilon =1$, then $\widetilde{\Lambda }=\Lambda >0$, and if $\epsilon =-1$, then $\widetilde{\Lambda }=-\Lambda <0$.
 
For the geometrical units $G=c=1$ the formula (1) takes the form in terms if $\epsilon $ and $\Lambda >0$
\begin{align}
S_{EP}=\frac{1}{32\pi}\int\limits_{\mathcal{D}}\Bigl (\eta _{ijkl}\vartheta^{i}\wedge \vartheta^{j}\wedge \Omega^{kl}+ \frac {\epsilon \Lambda}{6}\eta _{ijkl}\vartheta^{i}\wedge \vartheta^{j}\wedge \vartheta^{k}\wedge \vartheta^{l}\Bigr). 
\end{align}

Adding to the geometric part $S_{EP}$ the matter action 
\begin{align}
S_{m}=\int \limits_{\mathcal{D}}L_{mat}(\phi ^{A}, D\phi ^{A}, \vartheta^{i}),
\end{align}
where $\phi ^{A}$ means tensor-valued matter form and $D\phi ^{A}$ its absolute exterior derivative, we obtain full action
\begin{align}
S&=S_{EP}+S_{m} \notag
\\&=\frac{1}{32\pi}\int\limits_{\mathcal{D}}\Bigl (\eta _{ijkl}\vartheta^{i}\wedge \vartheta^{j}\wedge \Omega^{kl}+ \frac {\epsilon \Lambda}{6}\eta _{ijkl}\vartheta^{i}\wedge \vartheta^{j}\wedge \vartheta^{k}\wedge \vartheta^{l}\Bigr)+\int \limits_{\mathcal{D}}L_{mat}(\phi ^{A}, D\phi ^{A}, \vartheta^{i})
\end{align}

After some calculations one gets that the variation $\delta S=\delta S_{EP}+\delta S_{m}$ with respect to $\vartheta^{i},\ \omega^{i}_{\ j}$ and $\phi^{A}$ reads

\begin{align}
\delta S&=\int \limits_{\mathcal{D}}\Bigl [ \frac{1}{8\pi }\delta \vartheta^{i}\wedge \Bigl ( \frac{1}{2}\Omega ^{kl}\wedge \eta_{kli}+\epsilon \Lambda \eta_{i}+8\pi t_{i}\Bigr ) \notag
\\&+\frac{1}{2}\delta \omega ^{i}_{\ j}\wedge\Bigl (\frac{1}{8\pi }D\eta^{\ j}_{i}+s_{i}^{\ j}\Bigr )+\delta \phi^{A}\wedge L_{A}+ an \ exact \ form\Bigr ]. 
\end{align}

The three-forms: energy-momentum $t_{i}$, classical spin $s_{i}^{\ j}$ and $L^{A}$ are defined by the following form of the variation $\delta L_{m}$
\begin{align}
\delta L_{m}=\delta \vartheta^{i}\wedge t_{i}+\frac{1}{2}\delta \omega^{i}_{\ j}\wedge s^{\ j}_{i}+\delta \phi ^{A}\wedge L^{A}+ an \ exact \ form.
\end{align}
$\eta_{kli},\ \eta_{i}^{\ j}, \eta_{i}$ mean the forms introduced in the past by A. Trautman [11]. 

The variations $\delta\vartheta^{i}$, $\delta \omega^{i}_{\ j}$ and $\delta \phi^{A}$ are vanishing on the boundary $\partial \mathcal{D}$ of the compact domain $\mathcal{D}$.

Einstein's equations like all the other physical field equations arise due to variational principle, which is called the Principle of Stationary Action or Hamiltonian Principle. In our case it has the following form:
\begin{align}
\delta S=0, 
\end{align}

It leads us to the following sets of the field equations
\begin{align}
 \frac{1}{2}\Omega ^{kl}\wedge \eta_{kli}+\epsilon \Lambda \eta_{i}=-8\pi t_{i}
\end{align}
\begin{align}
D\eta^{\ j}_{i}=-8\pi s_{i}^{\ j} 
\end{align}
and
\begin{align}
L_{A}=0. 
\end{align}
$L_{A}=0$ represent equations of motion for matter field. These equations are not intrinsic in further our considerations, so we will omit them. We are interested only in the gravitational field equations which are given by the equations (8)-(9).

In vacuum where $t_{i}=s^{\ j}_{i}=0$ also $D\eta^{\ j}_{i}=0$ and we get standard vacuum Einstein's equations with cosmological constant $\widetilde{\Lambda }=\epsilon \Lambda $ 
\begin{align}
 \frac{1}{2}\Omega ^{kl}\wedge \eta_{kli}\pm \epsilon \Lambda \eta_{i}=0 
\end{align}
 and pseudoriemannian geometry.
 
In general, we have the Einstein-Cartan equations and Riemann-Cartan geometry (a metric geometry with torsion, see e.g. [11]).\\
The standard GR we obtain also if $\frac{\delta L_{m}}{\delta \omega ^{i}_{\ k}}=0 \implies s_{i}^{\ k}=0 \implies D\eta_{i}^{\ k}=0$, i.e., if we confine to spinless matter.

Namely, one has in the case the following gravitational equations
\begin{align}
\frac{1}{2}\Omega^{kl}\wedge \eta_{kli}+\epsilon \Lambda \eta_{i}=-8\pi t_{i}. 
\end{align}

One can show that $\frac{1}{2}\Omega^{kl}\wedge \eta_{kli}=-G_{i}^{s}\eta_{s}$, where the Einstein tensor $G_{i}^{s}$ is defined as follows
\begin{align}
G_{i}^{s}=R_{i}^{s}-\frac{1}{2}\delta_{i}^{s}R.
\end{align}

Putting $t_{i}=T^{\ s}_{i}\eta_{s}$ we get from (12)
\begin{align}
-G_{i}^{\ s}\eta_{s}+\epsilon \Lambda \delta_{i}^{s}\eta _{s}=-8\pi T_{i}^{\ s}\eta_{s}. 
\end{align}
or
\begin{align}
G_{i}^{\ s}-\epsilon \Lambda \delta_{i}^{s}=8\pi T_{i}^{\ s}.
\end{align}
(15) are standard Einstein equations with cosmological constant $\Lambda $ in tensorial notation with symmetric matter tensor: $T^{ik}=T^{ki}$.

\section{Einstein-Palatini action integral for General Relativity in vacuum and with nonzero cosmological constant $\widetilde{\Lambda }$ as integral action for a gauge field}

Now, getting back to Einstein-Palatini action in vacuum
\begin{align}
S_{EP}&=\frac{1}{4\kappa}\int \limits_{\mathcal{D}}\Bigl (\vartheta^{i}\wedge \vartheta^{j}\wedge \Omega^{kl}+ \frac{\epsilon \Lambda}{6}\vartheta^{i}\wedge \vartheta^{j}\wedge \vartheta^{k}\wedge \vartheta^{l}\Bigr )\eta_{ijkl} \notag \\
&=\frac{1}{4\kappa}\int \limits_{\mathcal{D}}\Bigl (\vartheta^{i}\wedge \vartheta^{j}\wedge \Omega^{kl}\eta_{ijkl}+ \frac{\epsilon \Lambda}{6}\vartheta^{i}\wedge \vartheta^{j}\wedge \vartheta^{k}\wedge \vartheta^{l}\eta_{ijkl}\Bigr ) 
\end{align}
and defining the duality operator $\star $ [1]
\begin{align}
\star :=-\frac{\eta_{ijkl}}{2} \quad \implies \eta_{ijkl}=-2\star 
\end{align}
one gets
\begin{align}
\eta_{ijkl}\Omega^{kl}=-2\star \Omega_{ij},
\\ \eta_{ijkl}\vartheta^{k}\wedge \vartheta^{l}=-2\star \bigl (\vartheta_{i}\wedge \vartheta_{j}\bigr ).
\end{align}
Thus the Einstein-Palatini action has the following form
\begin{align}
S_{EP}&=-\frac{1}{2\kappa}\int \limits_{\mathcal{D}}\Bigl [\vartheta^{i}\wedge \vartheta^{j}\wedge \star \Omega_{ij}+ \frac{\epsilon \Lambda}{6} \vartheta^{i}\wedge \vartheta^{j}\wedge \star \bigl (\vartheta_{i}\wedge \vartheta_{j}\bigr )\Bigr ]\notag\\
&=-\frac{1}{2\kappa}\int \limits_{\mathcal{D}}tr\Bigl [\vartheta \wedge \vartheta \wedge \star \Omega + \frac{\epsilon \Lambda}{6} \vartheta \wedge \vartheta \wedge \star\bigl (\vartheta \wedge \vartheta \bigr )\Bigr ].
\end{align}
Let us introduce the corrected curvature $\Omega_{cor}$
\begin{align}
\Omega_{cor}:=\Omega+\frac{\epsilon \Lambda}{3}\vartheta \wedge \vartheta \quad \implies \vartheta \wedge \vartheta =-\frac{3}{\epsilon \Lambda}\Bigl (\Omega-\Omega_{cor}\Bigr ). 
\end{align}
Substituting the last formula into Einstein-Palatini action we get 
\begin{align}
S_{EP}&=\frac{-1}{2\kappa}\int \limits_{\mathcal{D}}tr\Bigl [\vartheta \wedge \vartheta \wedge \star \Omega+\frac{\epsilon \Lambda}{6} \vartheta \wedge \vartheta \wedge \star\bigl (\vartheta \wedge \vartheta \bigr )\Bigr ]\notag \\
&=\frac{1}{2\kappa}\int \limits_{\mathcal{D}} tr\Bigl [\frac{3}{\epsilon \Lambda}\bigl (\Omega-\Omega_{cor}\bigr )\wedge\star \Omega -\frac{\epsilon \Lambda}{6}\frac{9}{\epsilon ^2\Lambda^{2}}\bigl (\Omega-\Omega_{cor}\bigr )\wedge\star \bigl (\Omega -\Omega_{cor}\bigr )\Bigr ] \notag \\
&=\frac{3}{4\Lambda \kappa \epsilon }\int \limits_{\mathcal{D}} tr\Bigl (2\bigl (\Omega-\Omega_{cor}\bigr )\wedge \star \Omega-\bigl (\Omega -\Omega_{cor}\bigr )\wedge \star \bigl (\Omega -\Omega_{cor}\bigr )\Bigr ) \notag \\
&=\frac{3}{4\Lambda \kappa \epsilon }\int \limits_{\mathcal{D}} tr\Bigl [2\Omega \wedge \star \Omega -2\Omega_{cor}\wedge \star \Omega -\Omega \wedge \star \Omega +\Omega_{cor}\wedge \star \Omega +\Omega \wedge \star \Omega_{cor}-\Omega_{cor}\wedge \star \Omega_{cor}\Bigr ] \notag \\
&=\frac{3}{4\Lambda \kappa \epsilon }\int \limits_{\mathcal{D}} tr\Bigl [\Omega \wedge \star \Omega -\Omega_{cor}\wedge \star \Omega +\Omega \wedge \star \Omega_{Cor}-\Omega_{cor}\wedge \star \Omega_{cor}\Bigr ] 
\end{align}
Because $-\Omega_{cor}\wedge \star \Omega +\Omega \wedge \star \Omega_{cor}$ reduces, then we finally have
\begin{align}
S_{EP}=\frac{3}{4\Lambda \kappa \epsilon }\int \limits_{\mathcal{D}} tr\Bigl [\Omega \wedge \star \Omega -\Omega_{cor}\wedge \star \Omega_{cor}\Bigr ].
\end{align}

The expression $tr\bigl (\Omega \wedge \star \Omega \bigr )=\eta_{ijkl} \Omega ^{ij}\wedge \Omega ^{kl}$ is in four dimensions a topological invariant called Euler's form, which does not influence the equations of motion [12]. Hence, in 4-dimensions the Einstein-Palatini action is equivalent to 

\begin{align}
S_{EP}=-\frac{3}{4\Lambda \kappa \epsilon }\int \limits_{\mathcal{D}} tr\Bigl (\Omega_{cor}\wedge \star\Omega_{cor}\Bigr ), 
\end{align}
where $\epsilon =\pm 1 $.

We see that the Einstein-Palatini action in 4-dimensions is efectively the functional which is quadratic function of the corrected Riemannian curvature, i.e., it has the form of the action for a gauge field.

Only one difference is that in (24) we have the star operator $\star $, which is different from Hodge star operator. Namely, our star operator acts onto "interior" indices (tetrad's indices), not onto forms as Hode duality operator does [2, 12].

It is interesting that $\Omega_{cor}=0$ for the de Sitter spacetime which is the fundamental vacuum solution to the Einstein equations (8) if $\epsilon =1$ and $\Omega_{cor}=0$ for the AdS spacetime if $\epsilon =-1$. The AdS spacetime is the fundamental solution of the equations (8) if $\widetilde {\Lambda }=\epsilon \Lambda <0$.

We would like to emphasize that in the case $\widetilde{\Lambda }=\epsilon \Lambda =0 \iff R=\infty$ the above trick with $\Omega_{cor}$ breaks. Namely, we have in this case $\Omega_{cor}=\Omega $. This result formally trivializes $S_{E-P}$ action (see formula (22)) to the strange form $S_{E-P}=0$ and has no physical meaning. The case $\widetilde{\Lambda }=\epsilon \Lambda<0$ needs introducing of the anti de Sitter spacetime (AdS) and its isometry group SO(3,2) (see Section 3). The anti de Sitter spacetime has very strange casual properties (see e.g. [18]). In consequence, it seems that the physical meaning of the case $\widetilde{\Lambda }=\epsilon \Lambda <0$ is problematic.

\newpage
\section{Geometrical interpretation of the corrected curvature $\Omega_{cor}$}

We begin from $\epsilon =1$, i.e., from $\widetilde{\Lambda }=\epsilon \Lambda=\Lambda >0$. This is de Sitter case because for $\widetilde{\Lambda }>0$, the Einstein equations (11) admit the de Sitter spacetime as fundamental solution (see,e.g. [18]). This spacetime is realized as hyperboloid 
\begin{align}
(\chi^{0})^{2}-(\chi^{1})^{2}-(\chi^{2})^{2}-(\chi^{3})^{2}-(\chi^{4})^{2}=-R^{2}
\end{align}
with radius $R$ in the 5-dimensional pseudoeuclidean spacetime $M(4,1)$ which possesses metric $\eta_{AB}=diag(1,-1,-1,-1,-1)$[18].

Let $P(M_{4}, GdS)$ denotes the principal bundle of de Sitter basis over a manifold $M_{4}$ (spacetime) with de Sitter group($ GdS$) [5, 13] as a structure group. The de Sitter group is isomorphic to the group $SO(4, 1)$ which acts on the spacetime $M(4,1)$ as rotations group.

Let $\widetilde{\omega}$ be 1-form of connection in the principle fibre  bundle $P(M_{4}, GdS)$. The form $\widetilde{\omega }$ has values in the algebra $\mathfrak{g}$ of the group $GdS$. The algebra $\mathfrak{g}$ is also the algebra of the group $SO(4,1)$. This algebra splits (as a vector space) into direct sum
\begin{align}
\mathfrak{g}=so(3, 1)\oplus R^{(3, 1)}.
\end{align}
Here $so(3, 1)$ denotes the algebra of the group SO(3, 1) isomorphic to Lorentz group $\mathcal {L}$, and $R^{(3, 1)}$ is a 4-dimensional vector space of generalised translations (translations in the curved de Sitter spacetime). One can identify the de Sitter spacetime with the quotient SO(4,1)/SO(3,1). \\
Let us define $so(3,1)=:\mathfrak{h}$, $R^{3,1}=:\mathfrak{p}$. Then we have [1,2]
\begin{align}
\mathfrak{g}=\mathfrak{h}\oplus \mathfrak{p},
\end{align}
and 
\begin{align}
[\mathfrak{h},\mathfrak{h}]\subset \mathfrak{h},\ [\mathfrak{h},\mathfrak{p}]\subset \mathfrak{p},\ [\mathfrak{p},\mathfrak{p}]\subset \mathfrak{h}.
\end{align}
This means that the Lie algebra $\mathfrak{g}$ is a symmetric Lie algebra [1,2].\\
On the other hand, the spaces which satisfy (27)-(28) are called globally symmetric Riemannian spaces [13].

Let $P(M_{4}, \mathcal {L})$  denotes the principal bundle of Lorentz basis over the manifold $M_{4}$. There exists a morphism of principal bundles 
\begin{align}
f:P(M_{4}, \mathcal {L}) \longrightarrow P(M_{4}, GdS)
\end{align}
analogical to the morphism of the bundle linear frames and the bundle affine frames [4]. This morphism is created by embedding of the $SO(3,1)$ group into $SO(4,1)$. It creates pull-back $f_{*}\widetilde{\omega }$ of the form $\widetilde{\omega}$ onto the bundle $P(M_{4}, \mathcal {L})$. Here $\widetilde{\omega}$ is the connection 1-form in the bundle $P(M_{4},GdS)$. \\
Let us denote this pull-back by $A$. $A$ is a 1-form on $P(M_{4}, \mathcal {L})$ with values in the direct sum [4]
\begin{align}
so(3, 1)\oplus R^{(3, 1)}.
\end{align}

Hence, we have a natural decomposition [2,3,4,13]
\begin{align}
A=f_{*}\widetilde{\omega}=\omega+\frac{\theta }{R} ,
\end{align}
where $\omega $ is a 1-form on $P(M_{4}, \mathcal {L})$ with values in the algebra $so(3, 1)$ and $\theta $ is a 1-form on $P(M_{4}, \mathcal {L})$ with values on $R^{(3, 1)}$. $\omega $ is a connection on the bundle $P(M_{4}, \mathcal {L})$. R is the radius of the de Sitter spacetime (see e.g. [18]).

On the base $M_{4}$ the 1-form $\theta $ can be identified with 1-form $\vartheta $ already used in this paper: $\theta=\vartheta $. In the following we will work on the base space $M_{4}$ and write (31) in the form 
$$A=\omega + \frac{\vartheta }{R}.$$

Let us compute the curvature 2-form $\widetilde{\Omega}$ of the pulled back $A$, where 
\begin{align}
A^{A}_{\ B}=\begin{cases}\   \ A^{a}_{\ b}=\omega ^{a}_{\ b} &\text {A,B=0,1,2,3,4}\\A^{a}_{\ 4}=\frac{1}{R}\vartheta ^{a}; A^{4}_{\ a}=\frac{1}{R}\vartheta _{a} &\text{a,b=0,1,2,3} \end{cases},
\end{align}
and $A^{AB}=-A^{BA}$.

From the definition we have
\begin{align}
\widetilde{\Omega}^{A}_{\ B}&=dA^{A}_{\ B}+A^{A}_{\ K}\wedge A^{K}_{\ B}.
\end{align}

Hence
\begin{align}
\widetilde{\Omega}^{a}_{\ b}&=dA^{a}_{\ b}+A^{a}_{\ k}\wedge A^{k}_{\ b}=d\omega^{a}_{\ b}+A^{a}_{\ d}\wedge A^{d}_{\ b}+A^{a}_{\ 4}\wedge A^{4}_{\ b}  \notag \\
&=\Omega ^{a}_{b}+\frac{1}{R}\vartheta ^{a}\wedge \frac{1}{R}\vartheta _{b}=\Omega ^{a}_{\ b}+\frac{1}{R^{2}}\vartheta ^{a}\wedge \vartheta _{b}
\end{align}

because in this case $A^{4}_{\ b}=\frac{1}{R}\vartheta _{b}$.
\begin{align}
\widetilde{\Omega}^{i}_{\ 4}&=dA^{i}_{\ 4}+A^{i}_{\ k}\wedge A^{k}_{\ 4}=\frac{1}{R}d\vartheta ^{i}+A^{i}_{\ b}\wedge A^{b}_{\ 4} \notag \\
&=\frac{1}{R}(d\vartheta ^{i}+\omega ^{i}_{\ b}\wedge \vartheta ^{b})=\frac{1}{R}\mathcal{D}_{\omega}\vartheta ^{i}=\frac{1}{R}\Theta^{i}_{\omega}.
\end{align}

In the last formula we have usual the antysymmetry of the connection form 
\begin{align}
A^{BC}=-A^{CB} \rightarrow A^{4}_{4}=A^{44}=0
\end{align}

(Indices A, B, C, ... are raised and lowered with the pseudoeuclidean metric $\eta _{AB}=\eta ^{AB}=diag (1,-1,-1,-1,-1)$ and the indices a, b, c, ... are raised and lowered with the metric $\eta_{ab}=\eta^{ab}=diag(1,-1,-1,-1)$).
So, we have obtained the final result
\begin{align}
\widetilde{\Omega}^{AB}=\begin{cases}\widetilde{\Omega}^{a}_{\ b}=\Omega _{\omega \ b}^{a}+\frac{1 }{R^{2}}\vartheta ^{a}\wedge \vartheta _{b}=\Omega _{\omega \ b}^{a}+\frac{\Lambda }{3}\vartheta ^{a}\wedge \vartheta _{b} \\\widetilde{\Omega}^{i}_{\ 4}=\frac{1}{R}\mathcal{D}_{\omega}\vartheta ^{i}=\frac{1}{R}\Theta ^{i}_{\omega } \end{cases}.
\end{align}
The cosmological constant $\Lambda =\frac{3}{R^{2}}>0$ and $\Theta $ means the torsion 3-form of the connection $\omega $.\\
In the Section 2 we gave the definition of the corrected curvature $\Omega_{cor}$ for the case $\epsilon \Lambda =\Lambda >0$ as follows
\begin{align}
\Omega_{cor}:=\Omega+\frac{\Lambda}{3}\vartheta \wedge \vartheta. 
\end{align}

As one can see this curvature is the so(3,1) part of the curvature $\widetilde{\Omega }$ of the connection $A=f_{*}\widetilde{\omega }=\omega +\vartheta $. If the torsion $\Theta $ of the connection $\omega $ is $0$, then $\Omega _{cor}=\widetilde{\Omega }$.

Let us consider now the case $\epsilon=-1$. Then $\widetilde{\Lambda }=\epsilon\Lambda =-\Lambda <0$. From the beginning we must remember that this case seems to have smaller physical meaning than the case $\widetilde{\Lambda }>0$. In the case $\widetilde{\Lambda }<0$ we have to take into account the principal bundle of the anti-de Sitter bases over spacetime manifold $M_{4}$. This principal bundle we will denote $P(M_{4}, AdS)$, where AdS means anti-de Sitter group. One can identify this group with the rotation group SO(3,2) in the pseudoeuclidean 5-dimentional spacetime $M^{3,2}$ with metric $G_{AB}=G^{AB}=diag(1,-1,-1,-1,1)$. On the other hand, the anti-de Sitter group is the isometry group of the anti-de Sitter spacetime (see e.g.[18]). The AdS spacetime is the fundamental solution to the Einstein equations (11) if $\epsilon =-1$, i.e., if $\widetilde{\Lambda }=-\Lambda <0$. This solution can be realized as 4-dimensional hyperboloid
\begin{align}
(\chi ^{0})^{2}-(\chi ^{1})^{2}-(\chi ^{2})^{2}-(\chi ^{3})^{2}+(\chi ^{4})^{2}=R^{2}
\end{align}
with imaginary radius $iR$ immersed in 5-dimensional spacetime $M(3,2)$ with metric $G_{AB}=diag(1,-1,-1,-1,1)$ (see e.g.[3,5]).

Let $\widehat{\omega }$ be 1-form of connection in the principal bundle $P(M_{4}, AdS)$. The form $\widehat{\omega }$ has values in the algebra $\widetilde{\mathfrak{g}}$ of the group SO(3,2). For the algebra $\widetilde{\mathfrak{g}}$ the formulas (27), (28) are correct.\\
Let us consider a morphism
\begin{align}
\widehat{f}:P(M_{4},\mathcal{L})\rightarrow P(M_{4},AdS)
\end{align}
generated by embedding Lorentz group $\mathcal{L}$ into SO(3,2) group. This morphism creates pull-back $\widehat{f}_{*}\widehat{\omega }$ of the form $\widehat{\omega }$ onto the bundle $P(M_{4},\mathcal{L})$.\\
Let us denote this pull-back by $\widehat{A}$.  $\widehat{A}$ is the 1-form on $P(M_{4},\mathcal{L})$ with values in the direct sum 
\begin{align}
so(3, 1)\oplus R^{(3, 1)}=\widetilde{\mathfrak{g}}.
\end{align}
Hence we have a natural decomposition [analogical to (31)]
\begin{align}
\widehat{A}=f_{*}\widehat{\omega }=\omega + \frac{\theta }{R}.
\end{align}
Here $\omega $ determines metric connection on the bundle $P(M_{4},\mathcal{L})$ and $\theta $ is a 1-form on $P(M_{4},\mathcal{L})$ with values in the space $R^{(3,1)}$ of the generalized translations in anti de Sitter spacetime. [$\theta $ is analogical to the soldering form on the bundle of the linear frames $P(M_{4}, GL)$]. R means the radius of the AdS spacetime. In this case one has $\widetilde{\Lambda }=\epsilon \Lambda =-\Lambda =\frac{-3}{R^{2}}$.\\
In the following we once more confine to the base manifold $M_{4}$ (=spacetime). Then, as in the case $\widetilde {\Lambda } >0$,
\begin{align}
\theta =\vartheta ,\ \ \ \widehat{A}=\omega +\frac{\vartheta }{R}. 
\end{align}
Let us calculate the curvature 2-form $\widehat{\Omega }$ of the pulled back $\widehat{A}$. Starting with
\begin{align}
A^{A}_{\ B}=\begin{cases}\   \ A^{a}_{\ b}=\omega ^{a}_{\ b} &\text {A,B=0,1,2,3,4}\\A^{a}_{\ 4}=\frac{1}{R}\vartheta ^{a}; A^{4}_{\ a}=\frac{1}{R}\vartheta _{a} &\text{a,b=0,1,2,3} \end{cases}.
\end{align}
and $A^{AB}=-A^{BA}$ we obtain, after calculations analogical to calculations performed in the case $\widetilde{\Lambda }=\epsilon \Lambda >0$ [Now, A, B, C, ... are raised and lowered with the metric $G_{AB}=G^{AB}=diag(1,-1,-1,-1,1)$]
\begin{align}
\widehat{\Omega}^{AB}=\begin{cases}\widehat{\Omega}^{a}_{\ b}=\Omega _{\ b}^{a}-\frac{1}{R^{2}}\vartheta ^{a}\wedge \vartheta _{b} \\\widehat{\Omega}^{i}_{\ 4}=\frac{1}{R}\Theta ^{i}_{\omega } \end{cases}
\end{align}
where $\Lambda =\frac{3}{R^{2}}>0$.

Here $\Omega $ is the curvature of the connection $\omega $ and $\Theta $ is its torsion. We see that in the case $\widetilde{\Lambda }=\epsilon \Lambda <0$, the $so(3,1)$ part of the curvature $\widetilde{\Omega }$ is equal 
\begin{align}
\Omega ^{a}_{\ b}-\frac{1}{R^{2}}\vartheta ^{a}\wedge \vartheta _{b}=\Omega ^{a}_{\ b}-\frac{\Lambda}{3}\vartheta ^{a}\wedge \vartheta _{b}
\end{align}
i.e., it is equal to $\Omega _{cor}$ given by (21) if $\epsilon =-1$. By using this $\Omega _{cor}$ one can easily obtain the form (24) (with $\epsilon =-1$) for the Einstein-Palatini action (16) with $\epsilon =-1$.

One can write the obtained results for $\widetilde{\Lambda }=\epsilon \Lambda \neq 0$, $\Lambda >0$, $\epsilon =\pm 1$ in the common form
\begin{align}
A^{A}_{\ B}=\begin{cases}\   \ A^{a}_{\ b}=\omega ^{a}_{\ b}\\A^{a}_{\ 4}=\frac{1}{R}\vartheta ^{a}; A^{4}_{\ a}=\frac{\epsilon }{R}\vartheta _{a}  \end{cases}.
\end{align}
\begin{align}
\widetilde{\widehat{\Omega}}^{AB}=\begin{cases}\Omega^{a}_{\ b}+\frac{\epsilon \Lambda }{3}\vartheta ^{a}\wedge \vartheta _{b} \\\Omega ^{i}_{\ 4}=\frac{1}{R}\Theta ^{i}, \ \Omega ^{4}_{a}=\frac{\epsilon}{R}\Theta _{a} \end{cases}.
\end{align}
where 
\begin{align}
\epsilon =\begin{cases}\   \ 1 &\text {for $\widetilde{\Lambda }>0$}\\-1 &\text{for $\widetilde{\Lambda }<0$} \end{cases}.
\end{align} 
In the Section 2 we gave the definition of the corrected curvature $\Omega _{cor}$ as follows:
\begin{align}
\Omega _{cor}:=\Omega+\frac{\epsilon \Lambda}{3}\vartheta \wedge \vartheta \ \ \ \ \ \Lambda >0. 
\end{align}
One can see that this curvature is a curvature of the connection pulled back from the bundles $P(M_{4},SO(4,1))$ or $P(M_{4},SO(3,2))$ onto bundle $P(M_{4},\mathcal{L})$
if $\Theta =0.$\\
If $\Theta \neq 0$ then $\Omega_{cor}$ is the $so(3,1)$-part of this curvature.

\section{Conclusion}

In this article we have shown that in four dimensions the action integral for GR with $\widetilde{\Lambda }\neq 0$ can be written in very similar form to the form of the action integral for the typical gauge field. There is only one difference - the star. Instead of the Hodge star, we have slightly different star called the duality operator [2, 12].\\
Our result is important because it shows that there is no need to generalize GR and construct very complicated gravitational theories to obtain a gravitational theory as a gauge theory. The ordinary GR formulated in terms of tetrads and spin connection with cosmological constant $\widetilde{\Lambda }\neq 0$ is already a gauge theory. The gauge group of this theory is Lorentz group $SO(3,1)$ or its double covering $SL(2,\mathbb{C})$. The above facts are very interesting in connection with universality of the Einstein theory (alternative theories are not necessary) [15,16] and in connection with trials of quantizing this theory (gauge field can be successfully quantized). 

Some scientists [1, 2, 3] were concerned with this problem and they came to the similar conclusions as ours, but they applied in their works the Cartan's approach to the connection in the principal bundle [2, 13, 14]. This approach is not well known among geometrists and relativists. We have used only the standard theory of connection in the principal bundle which was created by Ehresmann - Cartan's student [4, 8]. His approach is commonly used in differential geometry and in relativity.

\section*{Acknowledgements}
Authors would like to thank Prof. J. Ławrynowicz for possibility of delivering lecture at the Hypercomplex Seminar 2017.

\newpage
\section*{Appendix 1}
\begin{center}
\underline {$\eta$ forms and operations with them [11]}
\end{center}

Following [11] we define
\begin{align}
\eta _{ijkl}=\sqrt{|g|}\epsilon _{ijkl} \tag {A.1}
\end{align}
where $\epsilon _{ijkl}$ is Levi-Civita pseudotensor with properties
\begin{align}
\epsilon _{ijkl}=\begin{cases}\   \ 1 &\text {if the sequence of indices ijkl is an even permutation}\\ & \text{of the sequence 0, 1, 2, 3}; \\-1 &\text{if it is an odd permutation}; \\ \ \  0 &\text{if the sequence of indices ijkl is not an even permutation}\\ &\text{of the sequence 0, 1, 2, 3} \end{cases}.\tag {A.2}
\end{align}
and we take $\eta_{0123}=\sqrt{|g|}$. In Lorentzian coreper $|g|=1$.

One has [11]
\begin{align}
&\eta_{ijk}=\vartheta ^{l}\eta _{ijkl} \tag {A.3}
\\&\eta _{ij}=\frac{1}{2}\vartheta ^{k}\wedge \eta _{ijk} \tag {A.4}
\\&\eta _{i}=\frac{1}{3}\vartheta ^{j}\wedge \eta _{ij} \tag {A.5}
\\&\eta =\frac{1}{4}\vartheta ^{i}\wedge \eta _{i} \tag {A.6}
\\&\vartheta ^{n}\wedge \eta _{kli}=\delta ^{n}_{i}\eta _{kl}+\delta ^{n}_{l}\eta _{ik}+\delta ^{n}_{k}\eta _{li} \tag {A.7}
\\&\vartheta ^{m}\wedge \eta _{kl}=\delta ^{m}_{l}\eta _{k}-\delta ^{m}_{k}\eta _{l} \tag {A.8}
\\&\vartheta ^{j}\wedge \eta _{i}=\delta ^{j}_{i}\eta \tag {A.9}
\end{align}
The forms $\eta $,  $\eta_{i}$,  $\eta_{ij}$,  $\eta_{ijk}$ are Hodge dual to the forms $1$, $\vartheta^{i}$, $\vartheta^{i}\wedge \vartheta^{j}$, $\vartheta^{i}\wedge \vartheta^{j}\wedge \vartheta^{k}$ respectively [11].


\newpage
\begin{thebibliography}{10}
\addcontentsline{toc}{section}{References}
\bibitem{Wis}
D. K. Wise, {\em "MacDowell-Mansouri Gravuty and Cartan Geometry"}, CQG, \textbf{27} (2010) 155010 (arXiv:gr-qc/0611154v2, 15 May 2009)
\bibitem{Wise}
D. K. Wise, {\em "Symmetric Space Cartan Connections and Gravity in Three and Four Dimensions"}, arXiv:0904.1738v2 [math.DG], 3 August 2009
\bibitem{Randomo}
A. Randomo, {\em "Gauge Gravity: a forward-looking introduction"},   arXiv: 1010.5822v1 [gr-qc], 27 October 2010
\bibitem{Kobayashi}
S. Kobayashi, K. Nomizu, {\em "Foundations of Differential Geometry"}, Vol.1 and Vol.2,   Interscience Publishers, a division of John Wiley and Sons, New York ,  London 1963
\bibitem{Gursey}
F. G\"ursey, {\em "Introduction to Group Theory" an article in "Groups and Topology in Relativity"},C. DeWitt and B. DeWitt (editors), Gordon and Breach, London 1964
\bibitem{Dub}
A. Dubni\v ckova, {\em "Topological Groups for Physicists"}, Dubna 1987 (in Russian)
\bibitem{Moz}
J. Mozrzymas, {\em "Applications of Group Theory in Modern Physics"}, National Scientific Publishers PWN, Wrocław 1967 (in Polish)
\bibitem{Gan}
J. Gancarzewicz, {\em "Foundations of Modern Differential Geometry"}, SCRIPT, Warsaw 2010 (in Polish)
\bibitem{Sul}
R. Sulanke, P. Wintgen, {\em "Differentialgeometrie und Faserbuendel"}, Copyright by VEB Deutscher Verlag der Wissenschaften, Berlin 1972
\bibitem{Tra}
W. Kopczyński, A. Trautman, {\em "Spacetime and Gravitation"}, National Scientific Publishers PWN, Warszawa 1984 (in Polish - there exists English translation)
\bibitem{Tra1}
A. Trautman, {\em "Einstein-Cartan Theory"}, Symposia Mathematica, \textbf{12} (1973) 139
\bibitem{Tra2}
K. Hayashi, T. Shirafuji, {\em "Gravity from Poincare Gauge Theory of the Fundamental Particles. Part V" an article in "Progress of Theoretical Physics"}, \textbf{65} (1981) 525
\bibitem{Dre}
W. Drechsler, M.E. Mayer, {\em "Fiber Bundle Techniques in Gauge Theories" }, an article in ""Lectures Notes in Physics" Vol.67, Springer-Verlag, Berlin $\cdot $ Heidelberg $\cdot $ New York 1977
\bibitem{Sh}
R. W. Sharpe, {\em "Differential Geometry. Cartan's Generalization of Klein's Erlangen Program" }, Springer-Verlag, New York $\cdot $ Berlin $\cdot $ Heidelberg 2000
\bibitem{Tra}
J. Kijowski, {\em "International Journal of Geometric Methods in Modern Physics"}, \textbf13 (2016) 1640008
\bibitem{Tra1}
M. A. Schweizer, {\em "Gauge Theory and Gravitation"}, PhD, Zurich 1980
\bibitem{Tra2}
M. Dudek, J. Garecki, {\em "General Relativity with cosmological constant $\Lambda >0$ as a gauge theory"}, submitted to "International Journal of Geometric Methods in Modern Physics"
\bibitem{Tra3}
L. M. Sokołowski, {\em "Foundations of Tensor Analysis"}, Wydawnictwo Uniwersytetu Warszawskiego 2010.


\end{thebibliography}
\end{document}